\documentclass[aps,twocolumn,superscriptaddress]{revtex4}
\usepackage{amsmath}
\usepackage{amsfonts}
\usepackage{mathrsfs}
\usepackage{graphicx}
\usepackage{times}
\usepackage{color}
\usepackage{mathrsfs}

\def\be{\begin{equation}}
\def\ee{\end{equation}}
\def\bea{\begin{eqnarray}}
\def\eea{\end{eqnarray}}

\begin{document}

\title{Emergent interlayer nodal superfluidity of a dipolar fermi gas in bilayer optical lattices}

\author{Bo Liu}
\email{liubophy@gmail.com}
\affiliation{Department of Applied Physics, School of Science,
Xi'an  Jiaotong University, Xi'an  710049, Shaanxi, China}
\affiliation{Shaanxi Province Key Laboratory of Quantum Information and Quantum Optoelectronic Devices, Xi'an Jiaotong University, Xi'an 710049, Shaanxi, China}
\author{Peng Zhang}
\affiliation{Department of Applied Physics, School of Science,
Xi'an  Jiaotong University, Xi'an  710049, Shaanxi, China}
\author{Ren Zhang}
\affiliation{Department of Applied Physics, School of Science,
Xi'an  Jiaotong University, Xi'an  710049, Shaanxi, China}
\author{Hong Gao}
\affiliation{Department of Applied Physics, School of Science,
Xi'an  Jiaotong University, Xi'an  710049, Shaanxi, China}
\affiliation{Shaanxi Province Key Laboratory of Quantum Information and Quantum Optoelectronic Devices, Xi'an Jiaotong University, Xi'an 710049, Shaanxi, China}
\author{Fuli Li}
\affiliation{Department of Applied Physics, School of Science,
Xi'an  Jiaotong University, Xi'an  710049, Shaanxi, China}
\affiliation{Shaanxi Province Key Laboratory of Quantum Information and Quantum Optoelectronic Devices, Xi'an Jiaotong University, Xi'an 710049, Shaanxi, China}

\begin{abstract}
Understanding the interplay between magnetism and superconductivity is one of
the central issues in condensed matter physics. Such interplay induced nodal
structure of superconducting gap is widely believed to be a signature of exotic pairing mechanism
(not phonon mediated) to achieve unconventional superconductivity, such as in heavy fermion,
high $T_c$, and organic superconductors. Here we report a new mechanism to drive the interplay between magnetism and superfluidity via the spatially anisotropic interaction. This scheme frees up the usual requirement of suppressing long-range magnetic order to access unconventional superconductivity like through doping or adding pressure in solids.
Surprisingly, even for the half-filling case, such scheme can lead the coexistence of superfluidity and antiferromagnetism and interestingly an unexpected interlayer nodal superfluid emerges, which will be demonstrated through a cold atom system composed of a pseudospin-$1/2$ dipolar fermi gas in bilayer optical lattices. Our mechanism should pave an alternative way to unveil exotic pairing scheme resulting from the interplay between magnetism and superconductivity or superfluidity.
\end{abstract}

\maketitle

Searching for unconventional superconductors or superfluids and exploring their exotic pairing mechanism are some of the main
themes in condensed matter physics~\cite{2011_Norman_science}. In high temperature superconductors such as cuprates, heavy fermion intermetallic compounds and iron-pnictides, antiferromagnetism and superconductivity are two key phenomena~\cite{2006_Wen_RevModPhys,2004_Demler_RevModPhys,2001_Movshovich_PhysRevLett,2000_Hegger_PhysRevLett,2001_Sato_Nature,
2011_Stewart_RevModPhys}. Understanding the interplay between magnetism and superconductivity plays essential role for unveiling the unusual pairing mechanism of high temperature superconductivity, like spin-charge separation and RVB scenario~\cite{1987_Anderson_science,1988_Anderson_PhysRevB}. However, for most unconventional
superconductors, to unambiguously determine the detailed superconductivity mechanism like probing from the nodal gap structure is still an unresolved issue in solid state materials~\cite{2006_Matsuda_JournalofPhysics}. Besides the continuously growing efforts to study the interplay between magnetism and superconductivity in solids, there have been great interests of simulating unconventional superconductors and their exotic pairing mechanism in electronic systems via the cold atom based system in both experimental and theoretical studies, motivated by the recent experimental advances to create tunable interacting ultracold gases in optical lattices~\cite{2008_Bloch_RevModPhys,2008_Schneider_science,2008_Esslinger_nature,2011_Hemmerich_NatPhys,
2015_Hemmerich_PhysRevLett,2011_RevModPhys,2013_Galitski_nature,2014_Bo_arxiv,2016_Bo_PhysRevA,2016_Bo_PhysRevAI}. Such highly controllable atomic systems will not only provide a versatile tool for simulating
electronic systems, but also supply new probabilities to unveil new pairing scheme resulting from the interplay between magnetism and superfluidity with no counterpart in solids. One of the exciting experimental progress along this line
is the realization of antiferromagnetism in optical lattices~\cite{2015_Hulet_nature,2017_Greiner_nature,2011_Greiner_nature}. However, to further investigate the interplay between magnetism and superfluidity desires future experimental breakthroughs, in particular like to suppress heating and lower the temperature of the system~\cite{2010_Esslinger_Review}.

Here we report the discovery of a new mechanism to drive the interplay between antiferromagnetism and superfluidity.
The key idea here is to engineer the spatially anisotropic interaction through the special geometry of the system.
Surprisingly, the anisotropic interaction induced coexistence of superfluidity and antiferromagnetism can even occur for the half-filling case, dramatically distinguished from the previous studies like in heavy fermion or high $T_c$ superconductors where the superconductivity will not emerge without the suppression of long-rang magnetic order ~\cite{2007_Monthoux_nature,1998_Mathur_nature}. Furthermore, an unexpected nodal superfluid appears resulting from the interplay between antiferromagnetism and superfluidity here. To illustrate such an idea, we choose a specific cold atom system of a dipolar Fermi gas in bilayer optical lattices as an concrete example in this work. It is motivated by the recent rapid experimental progress in both magnetic dipolar atoms (such as $^{167}$Er~\cite{2014_Grimm_PRL,2014_Ferlaino_arxiv} and $^{161}$Dy~\cite{2012_Luming_PRL,2014_Benjamin_PRA} atoms) and polar molecules~\cite{2008_Junye_Science,2012_Martin_PRL}. It stimulates tremendous interests in exploring dipolar effects in many-body quantum phases~\cite{2012_Baranov_Reviews}. In particular, resulting from the anisotropic effect of dipole interaction, various exotic superfluids have been predicted in spinless or multicomponents dipolar fermi gas at low temperatures, such as a $p$-wave superfluid with the dominant $p_z$ symmetry, Weyl superfluidity~\cite{2015_Bo_PhysRevLett}, $p+ip$ superfluids in a single 2D plane~\cite{2009_Cooper_PRL,2012_Boliu_PRA,2008_Bruun_PRL} and the interlayer superfluidity in a bilayer or multilayer system~\cite{2010_Shlyapnikov_PhysRevLett,2012_wang_PhysRevA,2010_Potter_PhysRevLett}. As we shall show with the model below, the anisotropic dipole interaction induced interplay between antiferromagnetism and superfluidity can lead to other unexpected results.

\begin{figure}[t]
\begin{center}
\includegraphics[width=8cm]{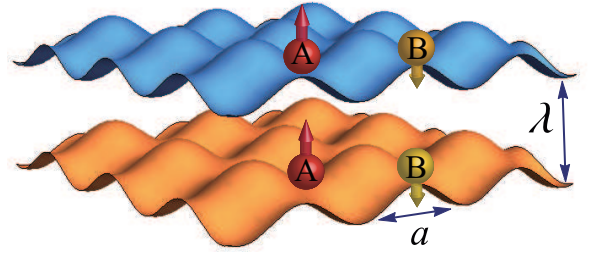}
\caption{Schematic picture of the bilayer optical lattice system. Within each single layer, there is a gas of interacting pseudospin-$1/2$ dipolar fermi atoms loaded in the same 2D square optical lattice with the lattice constant $a$. Here the
red and yellow balls attached with up and down arrows stand for the pseudospin up and down fermionic dipolar atoms, respectively. By assuming an external magnetic (electric) field applied, all the magnetic (electric) dipole moments are aligned along the same direction, which is perpendicular to the layers (the dipole moment is not shown in the figure). $\lambda$ is the interlayer distance. To treat with the intralayer antiferromagnetism, we decompose the square optical lattice within each single layer into two sublattices A and B, where the pseudospins are aligned oppositely between A and B site.} \label{figure1}
\end{center}
\end{figure}

\textit{Effective model $\raisebox{0.01mm}{---}$}
In order to design the spatially anisotropic interaction, which is the crucial ingredient in our new mechanism to drive the interplay between magnetism and superfluidity, let us consider a gas of interacting pseudospin-$1/2$ dipolar fermi atoms loaded in two parallel 2D optical lattices as shown in Fig.~\ref{figure1}. In both two layers, we consider the same lattice potential $V_{\rm OL}({\mathbf r})=-V[\cos^2(k_{L}x)+\cos^2(k_{L}y)]$, where $k_{L}$ is the wavevector of the laser field and the corresponding lattice constant is defined as $a=\pi/k_{L}$. Through applying an external magnetic (electric) field to align all the magnetic (electric) dipole moments along the same direction which is perpendicular to the layers, a specially anisotropic interaction among dipolar atoms can be engineered through such a bilayer geometric configuration. To be more specific, atoms in different layers attract each other at short range and repel each other at large distance, while within the same layer the interaction is purely repulsive.

A system of interacting pseudospin-$1/2$ dipolar fermions loaded in such a bilayer optical lattice can be described by the following Fermi-Hubbard model in the tight binding regime
\begin{eqnarray}
{H}&=& {H}_{intralayer}+{H}_{interlayer}
\label{hamiltonian}
\end{eqnarray}
and
\begin{eqnarray*}
{H}_{intralayer}&=&
-\sum_{\substack{<i,j> \\ \sigma, s}}t(c_{i\sigma s}^{\dag}c_{j\sigma s}+c_{j\sigma s}^{\dag}c_{i\sigma s})  \nonumber \\
&+&\frac{g}{2}\sum_{\substack{i, \sigma, \sigma^{\prime},s \\ \sigma \neq \sigma^{\prime}}}n_{i \sigma s}n_{i \sigma^{\prime} s} \nonumber
\end{eqnarray*}
\begin{eqnarray*}
{H}_{interlayer}&=&\frac{1}{2}\sum_{\substack{i, j, \sigma, \sigma^{\prime} \\ s, s^{\prime}, s \neq s^{\prime}}}V_{ij}c_{i\sigma s}^{\dag}c_{j\sigma^{\prime}s^{\prime}}^{\dag}c_{j\sigma^{\prime}s^{\prime}}
c_{i\sigma s}
\end{eqnarray*}
where $c_{i\sigma s}$ is the annihilation operator for the fermionic dipolar particle
at lattice site $\mathbf{R_i}$ in $s$ layer ($s=1,2$ labelling two layers).
$<i,j>$ denotes a summation over nearest neighbors in a single layer and
$\sigma(\sigma^{\prime})=\uparrow,\downarrow$ labels two species fermionic dipolar particles.
The onsite particle number operator for each layer is defined
as $n_{i \sigma s}=c_{i\sigma s}^{\dag}c_{i\sigma s}$ and $t$
describes hopping of fermions within one layer. Here we want to emphasize the crucial role
of the specially anisotropic interaction in our proposal. Through adjusting the external field
(e.g., dc electric field), the interaction between two dipolar atoms within one layer, i.e., intralayer
interaction, can be tuned to be dominated by the $s$-wave contribution~\cite{1998_Li_PhysRevLett,2001_Li_PhysRevA,2006_Ronen_PhysRevLett}. Therefore,
in the Hamiltonian Eq.~\eqref{hamiltonian}, the intralayer interacting strength is captured by
$g=g_{0}\int d\mathbf{x}\,|w^{i}_{\sigma s}|^{2}|w^{i}_{{\sigma^{\prime}} s}|^{2}$, where $g_{0}>0$ is determined by
the effective $s$-wave scattering length ~\cite{1998_Li_PhysRevLett} and $w^{i}_{\sigma s}$ is the Wannier function
at lattice site $\mathbf{R_i}$ in $s$ layer. Interestingly, due to the bilayer geometric configuration considered here,
the interaction between two dipoles belonging to different layers, i.e., interlayer interaction, has the form
\begin{eqnarray*}
V_{i-j}=d^2\frac{r_{ij}^{2}-2\lambda^2}{(r_{ij}^{2}+\lambda^2)^\frac{5}{2}}
\end{eqnarray*}
where $r_{ij}\equiv |\mathbf{R_i}-\mathbf{R_j}|$ is the in-plane separation between two dipoles, $\lambda$ is the interlayer spacing and $d$ is the dipole momentum. Such interlayer interaction is attractive for $r_{ij}<\sqrt{2}\lambda$, and repulsive at larger distance. In general, the intralayer repulsion is expected to lead the antiferromagnetic Mott state within each layer when considering half-filling, while the interlayer attraction should cause Cooper pairing instability between different layers which can coexist with the background antiferromagnetic state within each layer. Such a heuristically argued result is indeed confirmed by a self-consistent calculation through the model in Eq.~\eqref{hamiltonian}, to be introduced
below. It is not only strongly reminiscent of its counterpart in strongly correlated electronic materials, such as in  cuprate superconductor, but more importantly, the interplay between antiferromagnetism and superfluidity driven by the spatially anisotropic interaction will lead some unexpected properties to be illustrated in the following.

\textit{Antiferromagnetic Mott parent within each layer $\raisebox{0.01mm}{---}$}
Let us first discuss about the antiferromagnetic state within each layer. As we known, at half filling and for large repulsion, the model ${H}_{intralayer}$ describing dipolar fermions within single layer is expected to show antiferromagnetic order. Note that here we focus on the case with the stronger intralayer interaction compared to the interlayer counterpart, i.e., $g \gg |V_{i-j}|$, which allows us to start from the antiferromagnetic Mott parent in each single layer to further address the influence of the interlayer interaction. To study the intralayer antiferromagnetic state, we decompose the square optical lattice within each single layer into two sublattices A and B as shown in Fig.~\ref {figure1}, in such a way that all the neighbors of a site from sublattice A belong to sublattice B and the spins align oppositely at nearest neighbors between A and B sublattices. The antiferromagnetic order can be described by the following order parameter defined as $m_{s}\equiv<n_{i \uparrow s}-n_{i \downarrow s}>$ for
sublattice A and correspondingly $-m_{s}$ for sublattice B. So the Hamiltonian ${H}_{intralayer}$ in the momentum space describing the antiferromagnetic state within each single layer can be expressed as
\begin{eqnarray}
{H}_{intralayer}&=&\sum_{{\bf k}, \sigma, s}\varepsilon_{\bf k}(a_{{\bf k}\sigma s}^{\dag}b_{{\bf k}\sigma s}+b_{{\bf k}\sigma s}^{\dag}a_{{\bf k}\sigma s})\nonumber \\
&+&\frac{g}{2}\sum_{{\bf k}, s}[(1-m_s)(a_{{\bf k}\uparrow s}^{\dag}a_{{\bf k}\uparrow s}+b_{{\bf k}\downarrow s}^{\dag}b_{{\bf k}\downarrow s})\nonumber \\
&+&(1+m_s)(a_{{\bf k}\downarrow s}^{\dag}a_{{\bf k}\downarrow s}
+b_{{\bf k}\uparrow s}^{\dag}b_{{\bf k}\uparrow s})] \nonumber \\
&-&\sum_s\frac{N g m^2_s}{4}+\frac{N g}{2}
\label{hamiltonian 2}
\end{eqnarray}
where $a_{{\bf k}\sigma s}$ and $b_{{\bf k}\sigma s}$ are fermionic annihilation operators for sublattice A and B within $s$ layer and $N$ is the total number of lattice sites. ${\bf k}$-summation here is over the first Brillouin zone of a sublattice, and $\varepsilon_{\bf k}=-2t(\cos {k}_xa+\cos {k}_ya)$. We then introduce the quasi-particle operators which are given by
\begin{eqnarray}
\alpha_{{\bf k}\uparrow s}&=&u_{{\bf k}s}a_{{\bf k}\uparrow s}+v_{{\bf k} s}b_{{\bf k}\uparrow s}\nonumber \\
\alpha_{{\bf k}\downarrow s}&=&u_{{\bf k} s}b_{{\bf k}\downarrow s}+v_{{\bf k}s}a_{{\bf k}\downarrow s}\nonumber \\
\beta_{{\bf k}\uparrow s}&=&u_{{\bf k}s}b_{{\bf k}\uparrow s}-v_{{\bf k}s}a_{{\bf k}\uparrow s}\nonumber \\
\beta_{{\bf k}\downarrow s}&=&u_{{\bf k}s}a_{{\bf k}\downarrow s}-v_{{\bf k}s}b_{{\bf k}\downarrow s}
\label{transform}
\end{eqnarray}
where the coefficients are
\begin{eqnarray*}
u_{{\bf k} s}^2&=&1-v_{{\bf k} s}^2\nonumber \\
&=&\frac{1}{2}(1+\frac{g m_s}{2E_{{\bf k}s}})
\end{eqnarray*}
with $E_{{\bf k}s}=\sqrt{\varepsilon^2_{\bf k}+(\frac{g m_s}{2})^2}$ . So the Hamiltonian in Eq.~(\ref{hamiltonian 2})
can be diagonalized by a standard canonical transformation via the quasi-particle operators defined above
\begin{eqnarray}
{H}_{intralayer}&=&
\sum_{{\bf k}, \sigma, s}[(\frac{g}{2}-E_{{\bf k}s})\alpha_{{\bf k}\sigma s}^{\dag}\alpha_{{\bf k}\sigma s} \nonumber \\
&+&(\frac{g}{2}+E_{{\bf k}s})\beta_{{\bf k}\sigma s}^{\dag}\beta_{{\bf k}\sigma s}]\nonumber \\
&-&\sum_{s}\frac{Ngm^2_s}{4}+\frac{N g}{2}
\label{hamiltonian 3}
\end{eqnarray}
It is worth noting that the quasi-particles form four bands (doubly degenerated with respect to pseudospin) in each single layer with energies given by $E_{{\bf k}s}^{\pm}=\pm\sqrt{\varepsilon^2_{\bf k}+(\frac{g m_s}{2})^2}$. When considering half-filling case within each layer, it forms the antiferromagnetic Mott parent mimicking its counterpart in electronic materials such as cuprate, which is the starting point to discuss the superfluid instability between different layers as illustrated below.

\textit{Emergent interlayer nodal superfluidity $\raisebox{0.01mm}{---}$}
We now begin to investigate the interlayer interaction induced superfluidity arising from the antiferromagnetic
Mott parent within each single layer. As described above, in order to deal with the antiferomagnetism within each
layer, we divide the square optical lattice into two sublattices A and B. Therefore, it is natural to rewrite the
interlayer Hamiltonian ${H}_{interlayer}$ in terms of the operators defined on the sublattices as follows
\begin{eqnarray*}
{H}_{interlayer}&=&\sum_{\substack{{\bf k}, {\bf k}^{\prime}, {\bf q} \\ \sigma, \sigma^{\prime}, s, s^{\prime}\\ s \neq s^{\prime}}}\frac{V_{{\bf q}}}{N}(a_{{\bf k+q}\sigma s}^{\dag}a_{{\bf k}^{\prime}-{\bf q}\sigma^{\prime}s^{\prime}}^{\dag}a_{{\bf k}^{\prime}\sigma^{\prime}s^{\prime}}
a_{{\bf k}\sigma s}\nonumber \\
&+&a_{{\bf k+q}\sigma s}^{\dag}b_{{\bf k}^{\prime}-{\bf q}\sigma^{\prime}s^{\prime}}^{\dag}b_{{\bf k}^{\prime}\sigma^{\prime}s^{\prime}}
a_{{\bf k}\sigma s}\nonumber \\
&+&b_{{\bf k+q}\sigma s}^{\dag}a_{{\bf k}^{\prime}-{\bf q}\sigma^{\prime}s^{\prime}}^{\dag}
a_{{\bf k}^{\prime}\sigma^{\prime}s^{\prime}}
b_{{\bf k}\sigma s}\nonumber \\
&+&b_{{\bf k+q}\sigma s}^{\dag}b_{{\bf k}^{\prime}-{\bf q}\sigma^{\prime}s^{\prime}}^{\dag}b_{{\bf k}^{\prime}\sigma^{\prime}s^{\prime}}
b_{{\bf k}\sigma s})\nonumber \\
\end{eqnarray*}
where the momentum-summation is over the first Brillouin zone of the sublattice,
and $$V_{\bf q}= \sum_{m}V_{m}\exp(-i{\bf q}\cdot {\bf r}_m)$$ is the Fourier form of the interlayer
interaction $V_{i-j}$. Through the canonical transformation defined in Eq.~\eqref{transform}, the
interlayer Hamiltonian ${H}_{interlayer}$ can be further expressed via the quasi-particle operators.
From the analysis in the previous section, we know that the long-range antiferromagnetic
order makes the system to form four bands in each single layer. When considering half-filling case,
only the two lowest energy degenerated bands are filled with fermions and the other two are empty.
Therefore, to study the interlayer interaction induced superfluidity here, we can just consider the
Cooper pairs formed via the dipolar fermions in different layers both from the two lowest energy
degenerated bands of each single layer. Then the order parameter of the interlayer superfluid state
can be defined as
\begin{eqnarray}
\Delta_{\sigma^{\prime}\sigma}({\bf k})&=&\sum_{\nu {\bf k}^{\prime}}\frac{1}{N}f^{\nu}_{{\bf k} {\bf k}^{\prime}}V_{{\bf k} {\bf k}^{\prime}}<\alpha_{{\bf -k^{\prime}}\sigma^{\prime} s^{\prime}=2}\alpha_{{\bf k^{\prime}}\sigma s=1}> \nonumber \\
\label{gap}
\end{eqnarray}
where $<\ldots>$ means the expectation value in the ground state. The $k^{\prime}$-summation here
is over the first Brillouin zone of the sublattice and $\nu$-summation runs over the following coefficients
\begin{eqnarray}
f^{1}_{{\bf k} {\bf k}^{\prime}}&=&u_{{\bf k}s}u_{{\bf k}s^{\prime}}u_{{\bf k}^{\prime}s^{\prime}}u_{{\bf k}^{\prime}s}
+u_{{\bf k}s}v_{{\bf k}s^{\prime}}v_{{\bf k}^{\prime}s^{\prime}}u_{{\bf k}^{\prime}s}\nonumber\\
&+&v_{{\bf k}s}u_{{\bf k}s^{\prime}}u_{{\bf k}^{\prime}s^{\prime}}v_{{\bf k}^{\prime}s}
+v_{{\bf k}s}v_{{\bf k}s^{\prime}}v_{{\bf k}^{\prime}s^{\prime}}v_{{\bf k}^{\prime}s}\nonumber \\
f^{2}_{{\bf k}{\bf k}^{\prime}}&=&u_{{\bf k}s}u_{{\bf k}s^{\prime}}v_{{\bf k}^{\prime}s^{\prime}}v_{{\bf k}^{\prime}s}
+v_{{\bf k}s}v_{{\bf k}s^{\prime}}u_{{\bf k}^{\prime}s^{\prime}}u_{{\bf k}^{\prime}s}
\end{eqnarray}
Interestingly, from Eq.~\eqref{gap} we can find that the antiferromagnetic Mott parent within each single layer
will effectively modify the bare interlayer interaction $V_{i-j}$ through the term $f^{\nu}_{{\bf k} {\bf k}^{\prime}}$.
Surprisingly, it will lead a special nodal structure of the interlayer superfluid gap, which will be illustrated
below.

\begin{figure}[t]
\begin{center}
\includegraphics[width=8cm]{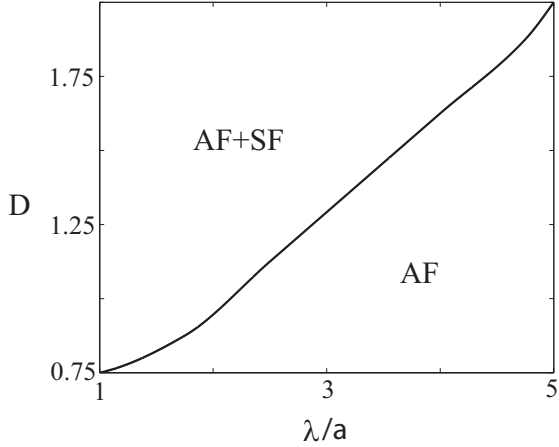}
\caption{Zero-temperature phase diagram as a function of the interlayer distance $\frac{\lambda}{a}$
and interlayer interaction strength $D=\frac{d^2}{t a^3}$ with a fixed intralayer interaction strength $\frac{g}{t}=20$
when considering half-filling case. There is a threshold of interlayer interaction strength marked by the solid line.
Blow that threshold, the antiferromagnetic Mott state is favored. While above that threshold, the interlayer superfluid
emerges and the system evolves into the coexistence of interlayer superfluidity and intralayer antiferromagnetism.} \label{figure2}
\end{center}
\end{figure}

We can also express the magnetic momentum $m_s$ within each single layer
determined by
\begin{eqnarray}
 m_s&=&\frac{2}{N}\sum_{\bf k}<a_{{\bf k}\uparrow s}^{\dag}a_{{\bf k}\uparrow s}-a_{{\bf k}\downarrow s}^{\dag}a_{{\bf k}\downarrow s}>
\label{magnetism}
\end{eqnarray}
in terms of the quasi-particle operators defined in Eq.~\eqref{transform}. Here $<\ldots>$ also means the expectation value in the ground state and the $k$-summation is over the first Brillouin zone
of the sublattice. Then the order parameter of the interlayer superfluid state $\Delta_{\sigma^{\prime}\sigma}$ defined in Eq.~\eqref{gap} together with the magnetic momentum $m_s$ constructed in Eq.~\eqref{magnetism} can be self-consistently
determined through the Bogoliubov transform to diagonalize the Hamiltonian in Eq.~\eqref{hamiltonian} under the mean-field
approximation.

\begin{figure}[t]
\begin{center}
\includegraphics[width=8cm]{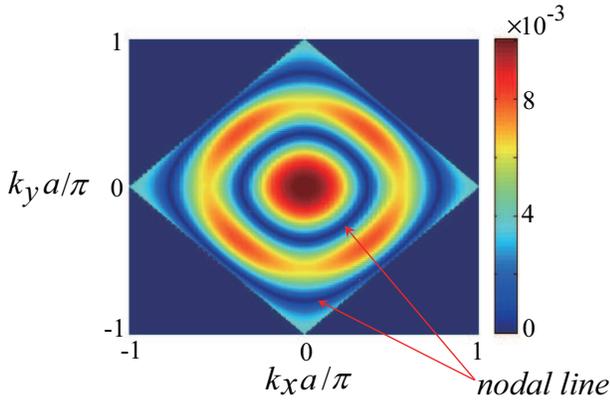}
\caption{The nodal gap structure of interlayer superfluid (showing the amplitude of superfluid gap $\Delta_{\uparrow\uparrow}$ here) resulting from the influence of the antiferromagnetic Mott parent within each single layer, where the interlayer distance $\frac{\lambda}{a}=1$, interlayer interaction strength $D=3$ and other parameters are the same as in Fig.~\ref{figure2}.} \label{figure3}
\end{center}
\end{figure}

Through numerically solving the Eq.~\eqref{magnetism} together with Eq.~\eqref{gap} self-consistently
to compute the ground state of Hamiltonian in Eq.~\eqref{hamiltonian}, the zero temperature phase diagram
of the system is obtained as shown in Fig.~\ref{figure2}. There are two different phases in the phase
diagram, which consists of an antiferromagnetic Mott state and the coexistence of interlayer superfluidity
and intralayer antiferromagnetism when considering half-filling case. As shown in Fig.~\ref{figure2}, there is a threshold strength of the interlayer
interaction characterized by $D=\frac{d^2}{t a^3}$ to support the emergence of the interlayer superfluidity.
Below that threshold, the intralayer repulsion will dominate and the ground state of the bilayer system is in
an antiferromagnetic state at zero temperature. While by increasing the interlayer interacting strength and
reducing the interlayer spacing, an interlayer superfluid in which Cooper pairs are formed via dipolar fermions from
different layers will occur to coexist with the antiferromagnetic Mott parent. The critical interlayer interaction
strength beyond which the interlayer superfluid emerges for different interlayer distance $\frac{\lambda}{a}$ is present in Fig.~\ref{figure2}, which is determined by monitoring the vanishing of the max value $\Delta_{max}\equiv \max(|\Delta_{\sigma^{\prime}\sigma}|)$ of superfluid order parameter. It is worth to note that an increase of interlayer interaction and a reduction of interlayer distance will make the emergence of the interlayer superfluid much easier.

Here we would like to emphasize some unique features of our new scheme to induce the interplay between antiferromagnetism and superfluidity via the spatially anisotropic interaction among dipolar atoms engineered from a bilayer geometric configuration.
First of all, distinguished from the strongly correlated electronic materials where it usually requires to suppress long-range magnetic order to access the coexistence of the antiferromagnetism and superconductivity
like through doping or adding pressure in cuprate, this new scheme frees up such a requirement. Second, even when the system is at half-filling, the coexistence of interlayer superfluidity and intralayer antiferromagnetism can also occur, that has no
counterpart in solids. Last but not the least, the intralayer antiferromagnetic Mott parent has a fantastic influence on
the interlayer superfluid. From Eq.~\eqref{gap}, we find that the background antiferromagnetic long-range order within each single layer effectively reforms the interlayer interaction. Surprisingly, it leads a special type of the nodal gap structure of interlayer superfluid as shown in Fig.~\ref{figure3}. It is strongly reminiscent of its counterpart in strongly correlated electronic materials. But more importantly, dramatically distinguished from solids like high $T_c$ superconductors, the nodal line here is not along the Fermi surface as illustrated in Fig.~\ref{figure3}, which characterizes  one of the unique features of our new scheme via using the spatially anisotropic interaction to induce the interplay between antiferromagnetism and superfluidity.

\textit{Conclusion $\raisebox{0.01mm}{---}$}
We have demonstrated a new approach to drive the interplay between antiferromagnetism and superfluidity via engineering
the spatially anisotropic interaction through considering special geometry of the system. We have shown that for the bilayer square optical lattice system with pseudospin-$1/2$ dipolar fermions,
even for the half-filled case within each single layer, the interlayer dipole-dipole interaction can lead the coexistence of the intralayer antiferromagnetism and interlayer superfluidity. Surprisingly, the background antiferromagnetic long-range order within each single layer leads a special type of the nodal gap structure of interlayer superfluid, which captures one of the new features of this scheme. This approach would complement with a new window in cold gases to realize and furthermore to control the interplay between magnetism and superfluidity.

\textit{Acknowledgment $\raisebox{0.01mm}{---}$} B. L. thanks helpful discussions with L. Yin and X. Li.
B. L. is supported by NSFC Grant No. 11774282.  P. Z. is
supported by NSFC Grant No. 11604255. H. G. is supported
by NSFC Grant No. 11774286. F. L. is supported
by NSFC Grant No. 11534008 and the National Key R\&D
Project (Grant No. 2016YFA0301404).

\bibliographystyle{apsrev}
\bibliography{bilayer}

\end{document}